# Oxygen-Induced Surface Reconstruction of SrRuO$_3$ and Its Effect on the BaTiO$_3$ Interface

Junsoo Shin,[†,‡,*] Albina Y. Borisevich,[‡] Vincent Meunier,[§] Jing Zhou,[⊥] E. Ward Plummer,[‖] Sergei V. Kalinin,[♦] and Arthur P. Baddorf,[♦,*]

[†]Department of Physics and Astronomy, The University of Tennessee, Knoxville, TN 37996, [‡]Materials Sciences and Technology Division, Oak Ridge National Laboratory, Oak Ridge, TN 37831, [§]Computer Science and Mathematics Division, Oak Ridge National Laboratory, Oak Ridge, TN 37831, [⊥]Department of Chemistry, University of Wyoming, Laramie, WY 82071, [‖]Department of Physics and Astronomy, Louisiana State University, Baton Rouge, LA, 70803, [♦]Center for Nanophase Materials Sciences, Oak Ridge National Laboratory, Oak Ridge, TN 37831.

## ABSTRACT

**Atomically engineered oxide multilayers and superlattices display unique properties responsive to the electronic and atomic structures of the interfaces. We have followed the growth of ferroelectric BaTiO$_3$ on SrRuO$_3$ electrode with *in situ* atomic scale analysis of the surface structure at each stage. An oxygen-induced surface reconstruction of SrRuO$_3$ leads to formation of SrO rows spaced at twice the bulk periodicity. This reconstruction modifies the structure of the first BaTiO$_3$ layers grown subsequently, including intermixing observed with cross-section spectroscopy. These observations reveal that this common oxide interface is much more interesting than previously reported, and provide a paradigm for oxygen engineering of oxide structure at an interface.**

[*] Address correspondence to jshin@ornl.gov, baddorfap@ornl.gov.

Advances in atomically-controlled oxide growth have generated new classes of materials with unique physical properties highly sensitive to abrupt interfaces.[1–4] The extreme sensitivity of oxides to electron concentration is coupled to charge transfer, structure, and spin to produce spectacular behavior including interface mediated conduction,[3,5,6] superconductivity,[7] magnetism,[8] and phase transitions[9,10] in parent materials lacking these attributes. The response of oxides to structural instabilities or disorder is greater in two-dimensional systems, seen in examples of Anderson localization, Peierls instability or charge density wave transitions. The central role of oxygen stoichiometry has been repeatedly shown in defining both structure and properties of oxide interfaces,[11–13] and can be tuned as a means to control static and dynamic distributions of electrons and atoms for a new generation of functional materials with applications ranging from oxide sensors and electronics to energy capture and storage. Nevertheless, few atomic scale studies of interface structures exist for complex oxides, due to a need for multiple tools to probe subsurface features, the need for a highly controlled environment,[14,15] and the insulating nature of many oxides.

We have studied the structural evolution of surfaces and interfaces during the layer-by-layer growth of $BaTiO_3$ films on $SrRuO_3$. This pair combines the classic ferroelectric, $BaTiO_3$, and the most common conducting oxide, $SrRuO_3$, and has been the subject of a number of investigations.[11,16] By combining *in situ* measurements of in-plane surface structure, *ex situ* cross sectional microscopy and spectroscopy, and first principles simulations, we observed the atomic structure of $SrRuO_3$ surfaces and its impact on the interface and structure of several layers of $BaTiO_3$. Surprisingly, the $SrRuO_3$ surface, which has conventionally been considered flat based on observation of well-separated, single layer steps in *ex situ* atomic force microscopy (AFM), is reconstructed at atomic length scales. This reconstruction increases the oxygen concentration and leads to both intermixing and structural change in $BaTiO_3$ at the interface. Clearly, AFM lacking atomic resolution cannot be relied upon to identify structural or stoichiometric deviations. *In situ* characterization of films must become the norm to identify the fundamental origins of behavior.

$SrRuO_3$ and $BaTiO_3$ films were grown on (001) oriented $SrTiO_3$ substrates using pulsed laser deposition with protocols described in the Methods section. To study the interface structure between $BaTiO_3$ and $SrRuO_3$ films, 1, 2, 4 and 10 unit cell $BaTiO_3$ films were grown on $SrRuO_3/SrTiO_3$. The detailed deposition conditions and the *in situ* electron diffraction experiments of $BaTiO_3/SrRuO_3$ films on $SrTiO_3$ are reported elsewhere.[17]

The growth quality and quantity of deposited material were confirmed using Reflection High Energy Electron Diffraction (RHEED). As shown in Figure 1A and Figure 1B, diffraction intensities oscillated as materials were deposited with each oscillation indicating one layer of film growth. For $SrRuO_3$, RHEED intensities oscillated several times but quickly reached a steady state, which has been shown to reflect a transition from layer-by-layer growth to step-flow growth with a SrO surface termination.[18] $BaTiO_3$ growth produced extended oscillations indicating good layer-by-layer growth and revealing the number of layers (one per oscillation) as they were grown. The flatness of these films was assured by removing several samples from vacuum for Atomic Force Microscopy (AFM). While exposure to atmosphere results in adsorption which could affect the surface structure,[14] ambient AFM images were useful as a comparison to many previous studies. Results shown in Figures 1C, D, and E, show these films were smooth with only single unit-cell steps and step densities similar to that of the $TiO_2$-terminated $SrTiO_3$ substrates.

The expectation, then, was that the interface between SrRuO$_3$ and BaTiO$_3$ would be atomically flat. Buried interfaces are notoriously difficult to characterize on an atomic scale; scanning probes, or other surface methods, cannot access the region of interest. Here, we examined this interface with cross-sectional Scanning Transmission Electron Microscopy (STEM). In Z-contrast STEM imaging, the intensity of an atomic column in the image is roughly proportional to the square of its atomic number, providing contrast between the two materials. The interface between SrRuO$_3$ and BaTiO$_3$ is clearly seen in Figure 2A where TiO$_2$ planes take over from more intense RuO$_2$ planes. Image profiles help quantify the transition. The profile in Figure 2B corresponds to the box on the image in Figure 2A, *i.e.* it represents an average over 6 atomic rows parallel to the interface. A general intensity decrease from left to right in the image originates from the decreasing specimen thickness. The individual column intensities follow the composition, for example, the SrO termination of the SrRuO$_3$ is clearly observed. Interestingly, the first BaO column marked Ba* has a considerably reduced intensity relative to others. This suggests the depletion of Ba or presence of Sr in this column. If, however, we construct a profile from individual rows across the image and track the intensities of the SrO and BaO columns closest to the interface, it becomes clear that this compositional change is not uniform. In Figure 2C we plot the corresponding peak heights (obtained from Gaussian fits of the profiles) in the image in Figure 2A as a function of vertical coordinate (along the interface). The last SrO layer and the first BaO layer have variable profiles, implying changing compositions. It is important to discern that these layers are correlated in composition, *i.e.* the two layers nearest the interface can be both BaO, both SrO, or both mixed at about the same degree. The composition of these layers varies on a nanometer scale forming "domains" along the interface. The presence of mixed BaO/SrO columns does not necessarily imply mixing on an atomic scale; because the observed compositional domains could have dimensions as small as 2 nm (see Figure 2C), there could be overlap in the beam direction (normal to the image plane) resulting in apparent mixing. The cumulative BaO/SrO ratio in these rows calculated over the entire image is very close to 1:1, suggesting that stoichiometry is preserved overall. The observed contrast pattern indicates that the interface has a complex structure, possibly similar to the schematic in Figure 2D, but not atomically smooth.

These observations led us to examine the $SrRuO_3$ surface in more detail. Higher resolution, obtained with *in situ* scanning tunneling microscopy (STM) revealed a more complex picture of the $SrRuO_3$ surface. The images in Figure 3 were acquired at 1.4V and 40 pA. STM images, such as those shown in Figures 3A and B, resolve features the size of an individual unit cell (0.4 nm). These images show rows of atoms in the surface plane oriented along (110) and (1-10) crystallographic directions. These rows are interspersed with a number of missing atoms, creating a large fraction of defects in the ordering. Line analysis (Figure 3C) of the STM images show that the spacing along the rows is 0.6±0.05 nm and corrugation about 0.004±0.002 nm while the spacing between the rows (Figure 3D) is 1.2±0.05 nm with much larger corrugation of 0.10±0.01 nm. Along the rows, therefore, the spacing corresponds to a single unit cell ($\sqrt{2}$ times the lattice constant of $SrTiO_3$, 0.391 nm), while the spacing between rows is twice that value, *i.e.* every second substrate unit cell. It is clear that these rows and holes, unobserved with ex situ AFM, will provide a profound effect on the surface and interface properties and on subsequent growth mechanisms of materials such as $BaTiO_3$.

Low Energy Electron Diffraction (LEED) revealed the evolution of long-range ordered structures at several stages of the film growth. As this technique involves scattering of electrons with energies typically between 50 and 200 eV, a conducting substrate is needed to avoid charging which would mask the diffraction pattern. Patterns were obtained from $SrRuO_3$ and from thin films of $BaTiO_3$ on $SrRuO_3$. As shown in Figure 5B, the diffraction pattern from $SrRuO_3$ thin film surfaces had not only the square pattern expected from a bulk-terminated film, but also an addition spot halfway between each bulk spot. This pattern showed the surface unit cell periodicity was doubled in both surface crystallographic directions, a periodicity known as p(2x2). This diffraction pattern is consistent with the rows observed in STM that were separated by twice the bulk lattice constant. However, the STM images revealed that the local periodicity is established by rows along either (100) or (010) directions, *i.e.* local domains of (2x1) and (1x2) symmetry that sum together to appear as a p(2x2) pattern.

Analysis of the growth of $SrRuO_3$ on STO has shown that the surface is terminated by a SrO layer, with $RuO_2$ below.[18] Since Sr has a greater density of conducting electronic states than oxygen, Sr atoms most likely are imaged by STM. The images suggested therefore that Sr or a Sr oxide is responsible for the rows. The holes between rows (*i.e.* where rows are incomplete) are similar to those seen with STM on surfaces of layered Sr-Ru oxides, including $Sr_2RuO_4$ and $Sr_3Ru_2O_7$, although these materials exhibit a c(2x2) symmetry, without extended rows.[19] This corrugated, imperfect surface can also help explain the unexpected surface reactivity of the surface when exposed to atmosphere.[14,20]

To identify the rowed structure observed experimentally, we examined several structural candidates using first principles density functional theory (DFT). We initially verified the effect of removing single SrO dimers from the SrRuO$_3$ surface; a dimer being charge neutral is more likely than a single atom to produce a stable structure. The observed vertical corrugation between rows in STM was 0.1 nm, not too different than the ~ 0.15 nm dip corresponding to the rigid removal of a pair from bulk SrRuO$_3$. However, the computed energy to remove a SrO dimer is prohibitively large, costing ~ 7 eV/pair. This large extraction energy can be understood since the process involves breaking covalent bonds with an accompanying energy penalty that is not counterbalanced by the creation of other bonds.

We next investigated the energetics related to the removal and "elevation" of a SrO dimer onto the surface (*i.e.* the SrO pair is promoted from the surface layer over the surface), in such a way that part of the energy associated with the creation of the SrO vacancy is compensated by bond formation with top atoms. This could be expected to be an unstable configuration, and indeed, during the calculation most of the initial configurations relax back in to the cleaned, defect-free surface (Figure 4A). However, appropriate surface structures, for example the Sr and O geometry of Figure 4B, did support bonding creating a local energy minimum, *i.e.* a metastable configuration.

In the observed (2x1)+(1x2) rows, isolated defects appear less stable than a row of defects. Computationally, we found that a row of SrO is more stable, by 0.32 eV/pair, and that the preferential ordering was along the (110), or equivalently (1-10), direction. The structure corresponding to complete rows of defects is shown in Figure 4B, and corresponds to a formation energy of +5.29 eV/pair when compared to a defect-free surface. As expected the defect formation energy was significantly lowered when the SrO remains bound to the surface, rather than simply ejected from it. Nevertheless, the reduced number of Ru-O and Sr-O bonds still leaves the energy cost too high to explain the observations.

In the oxygen rich atmosphere required to approach stoichiometric growth of oxides such as SrRuO$_3$, molecular oxygen should be very reactive with displaced Sr as described above. We focused next our attention on O interactions with the rowed SrO structure, where SrO is promoted onto the surface. In Figure 4C, we present the minimum energy structures resulting from the interaction of O$_2$ with the system of Figure 4B. When a single O is added per displaced SrO, its stable position is directly above the underbonded Ru (small white sphere), close to the position of the displaced O in the pristine surface. More importantly, the system energy is considerably reduced by 4.76 eV/O, which is much larger that the corresponding 1.2 eV/O adsorption energy on a defect-free surface. In other words, the defect energy was reduced to 1.78 eV/SrO. This formation energy, calculated at 0 K, is sufficiently low to suggest that SrO rows, displaced from the surface layer, together with excess oxygen, could produce the observed structure. The calculations continue to indicate that this structure should be metastable, which is consistent with STM measurements that will be discussed elsewhere.

As 1-2 layers of BaTiO$_3$ were grown on SrRuO$_3$, the LEED pattern remained p(2x2) (Figure 5B), however the relative intensities of diffraction spots were altered from those observed from SrRuO$_3$ alone. This change in relative intensities indicates a change of structure, with two important implications. First, this pattern must represent the order of the BaTiO$_3$ film, and cannot arise solely from exposed remnants of SrRuO$_3$. Second, this shows that the SrRuO$_3$ reconstruction influences the structure of the BaTiO$_3$ at the interface, which does not share the symmetry of bulk BaTiO$_3$, but instead has a periodicity two times larger in the plane of the interface.

Growth of thicker BaTiO$_3$ reverts the pattern observed in LEED to the (1x1) symmetry of the bulk. As shown in Figure 5B, as few as 4 layers of BaTiO$_3$ produce at a (1x1) periodicity; the additional diffraction spots indicating a doubled unit cell are gone. STM images of Figure 5C and D show the local periodic atomic rows along either (100) or (010) direction whereas Figure 5E no longer reveal rows. The same periodicity and surface topography remain in films of 10 layers of BaTiO$_3$. The LEED technique is highly surface sensitive, owing to the short mean-free path of low energy electrons in matter. Consequently, while the surfaces of these films show no reconstruction, *i.e.* no deviation of the in-plane symmetry from the bulk, the p(2x2) structure could persist at the interface. A reconstructed interface would have a profound influence on our understanding and modeling of phenomena such as the recent studies showing well-defined interfaces with remanent polarization down to 3.5 nm,[21,22] interface closure domains,[23] and ferroelectric effects on electron tunneling[24] in ultrathin films.

It is clear from these combined studies using STM, STEM, LEED, and DFT theory that the surfaces and interfaces of oxides as common as SrRuO$_3$ and BaTiO$_3$, each widely applied for their conductivity and ferroelectricity, respectively, can be more complex than previously assumed. The goal of an atomically abrupt interface in heteroepitaxy can be foiled by the intrinsic differences of terminated materials. In this example, conventional AFM imaging failed to identify a restructuring of the SrRuO$_3$ surface, where excess oxygen leads to rows of SrO along (100) or (010) crystallographic directions that double the unit cell periodicity. When buried under a BaTiO$_3$ film, these rows lead to an interface with mixed SrO and BaO composition and an in-plane doubling of the periodicity of the first few BaTiO$_3$ layers. This combined approach, including *in situ* characterization and modeling at an atomic scale, presents a new archetype for identification of oxygen stoichiometry and interface structure required for control of functional properties, for example ferroelectricity and transport in thin films.

## METHODS

A 15-nm-thick SrRuO$_3$ film was deposited on TiO$_2$-terminated[25] SrTiO$_3$ (001) substrate by PLD with layer thickness and growth mode monitored by high-pressure RHEED in the growth chamber with a base pressure 1x10$^{-8}$ Torr. Growth parameters of substrate temperature and oxygen pressure were 700 °C and 100 mTorr with an average deposition flux of 0.05 unit cells/s.[26] A KrF excimer laser ($\lambda$ = 248 nm) was used for growing films at a repetition rate of 5 Hz. After deposition, samples were kept *in situ* and annealed at 450 °C for 90 min in 1 Torr O$_2$ after growth and then cooled down to room temperature. Subsequently, the pressure was lowered and the samples were transferred without exposure in air to the ultra high vacuum (2x10$^{-10}$ Torr) STM and LEED chambers.

Calculations for structural investigation were performed within DFT, using the Vienna ab initio simulation package (VASP).[27,28] The Kohn-Sham equations were solved using the projector augmented wave (PAW) approach[29,30] and a plane-wave basis with a 400 eV energy cutoff and the exchange-correlation functional was represented by the Local Density Approximation (LDA).[31] Spin polarized calculations were used throughout. The system was set-up as follows: first we relaxed a SrTiO$_3$ (STO) unit cell in bulk using a 12x12x12 Monkhorst-Pack Brillouin zone sampling, resulting to a crystal structure with a=b=0.546 nm and c=3.863 nm. We used a single 2x2 slab of STO along as support. The atoms in this slab were kept fixed during the course of all the simulations. Imposing the lattice constants of STO in bulk, this results in a unit cell of 1.092 x 1.092 nm2 for the planar dimension. The system, called cleaned or defect-free surface hereafter, was obtained by adding and relaxing a 2x2 (001) slab of SrRuO$_3$ on top of the previously position STO, resulting in a 96 atom unit cell. We chose the c-axis of the working unit in such a way as to ensure a minimum of 0.7 nm of vacuum between periodic images. That unit cell was used as a starting point for all the calculations shown here. Note that we used a 4x4x1k-point grid for the slab calculations.

## *ACKNOWLEDGMENTS*

Research was sponsored by the Division of Materials Science and Engineering (JS) and at the Center for Nanophase Materials Sciences (SVK and APB) by the Scientific User Facilities Division, at Oak Ridge National Laboratory, for the Office of Basic Energy Sciences, U. S. Department of Energy. This work was also partially supported by the Department of Energy Grant DE-SC0002136 (EWP).

**Figure Captions**

**Fig. 1.** *In situ* RHEED oscillation during and RHEED pattern after (A) SrRuO$_3$ deposition and (B) 10 unit cells of BaTiO$_3$ deposition. *Ex situ* AFM topography of 3x3 μm$^2$ regions of (C) 15 nm thick SrRuO$_3$ thin film, (D) 2 unit cells thick BaTiO$_3$, and (E) 10 unit cells thick BaTiO$_3$ thin films on SrRuO$_3$/SrTiO$_3$.

**Fig. 2.** (A) High angle annular dark field STEM image of the BaTiO$_3$/SrRuO$_3$/SrTiO$_3$ film in cross-section; (B) intensity profile averaged over area in the blue box in (A), showing decreased intensity in the first Ba column denoted as Ba*, (C) Ba and Sr column intensities in the vicinity of the interface (color coding on right) as a function of vertical coordinate of the image (black Ba and green Sr correspond to the first columns near the interface), (D) interface structure suggested from STEM images, with ideal interface position indicated by red line (Sr – green, Ru – grey, O- red, Ba – purple, Ti – blue).

**Fig. 3.** Scanning Tunneling Microscopy images of the thin film SrRuO$_3$ surface showing (A) rows and their long range periodicity, (B) a higher resolution scan with line profiles, (C) along the rows where the spacing is 0.6 nm, and (D) across the rows which are separated by 1.2 nm. The (2x1) unit cell (white rectangle) is indicated in (B).

**Fig. 4.** SrRuO$_3$ models viewed from the top (upper panels) and side (lower panels). (A) The bulk terminated, flat surface model. (B) The minimum energy configuration when one row of SrO is "elevated" on top of the surface. This structure is not stable. (C) Addition of one O per

promoted SrO, which energetically prefers the site vacated by the Sr, increases the stability substantially.

**Fig. 5.** Tunneling and transmission microscopy images and electron diffraction combined to show the development of the structure of $BaTiO_3$ on $SrRuO_3$ on a $SrTiO_3$ substrate. (A) The center image shows a cross-section acquired with z-contrast STEM. (B) The upper three images are LEED diffraction patterns at the indicated stage of growth taken at 70 eV for the $SrRuO_3$ and at 180 eV for the $BaTiO_3$ surfaces; the first two show a p(2x2) unit cell doubling, while the third is bulk-like (1x1). (C), (D), and (E) STM images of the $SrRuO_3$ surface with atomically resolved rows, and $BaTiO_3$ surfaces from 2 and 10 layer films.

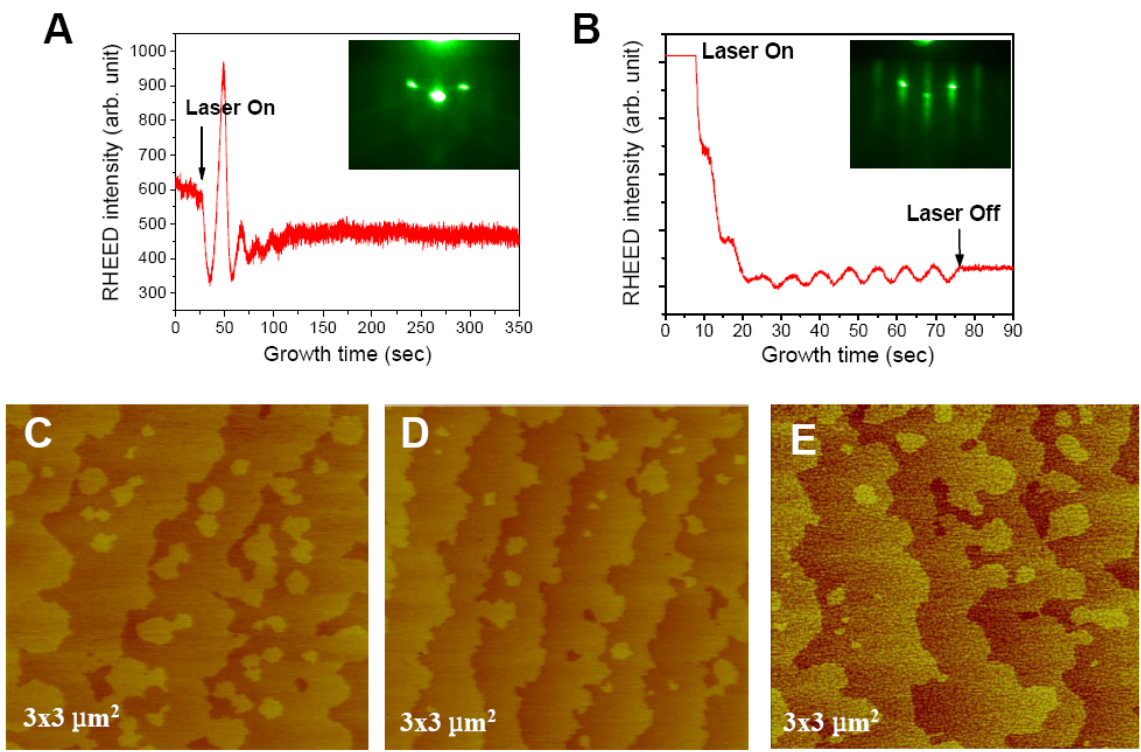

**Figure 1**

**Figure 2**

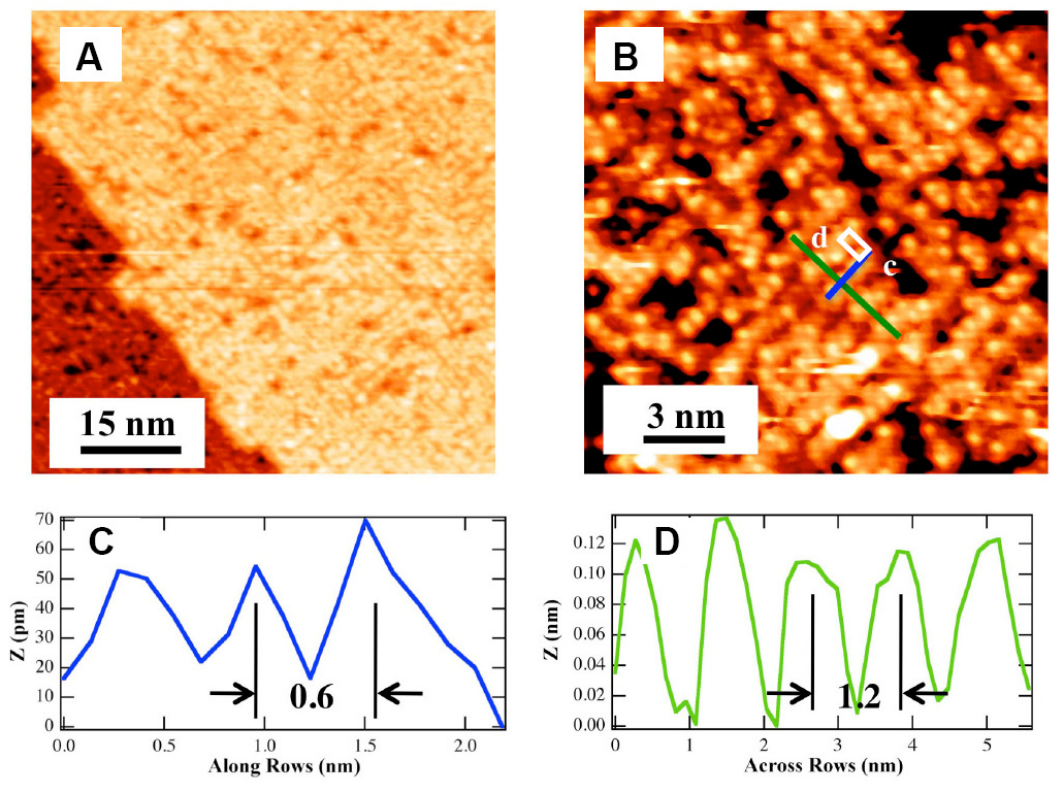

**Figure 3**

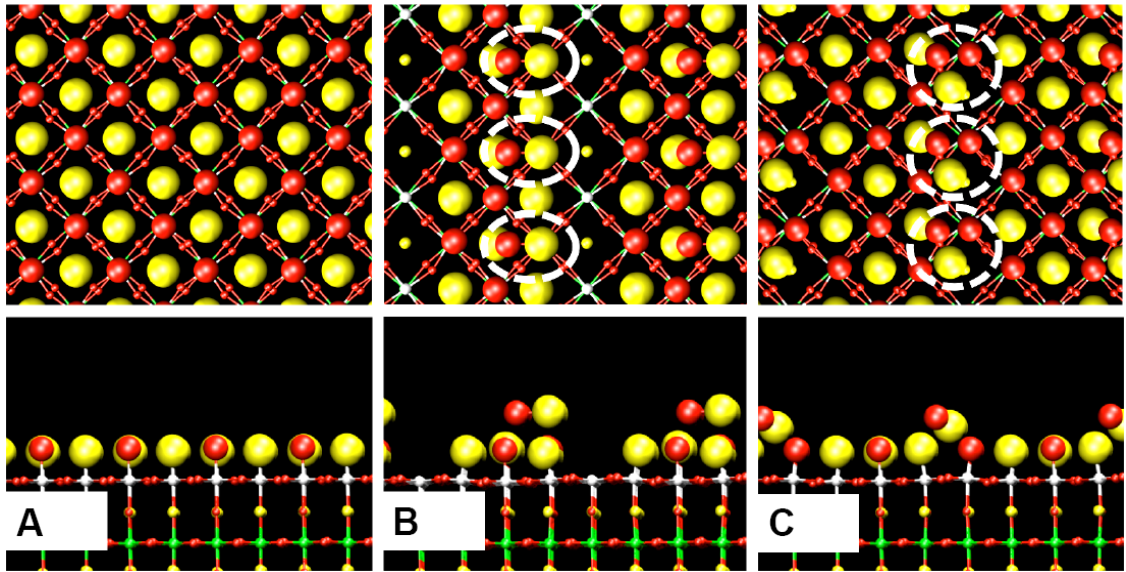

**Figure 4**

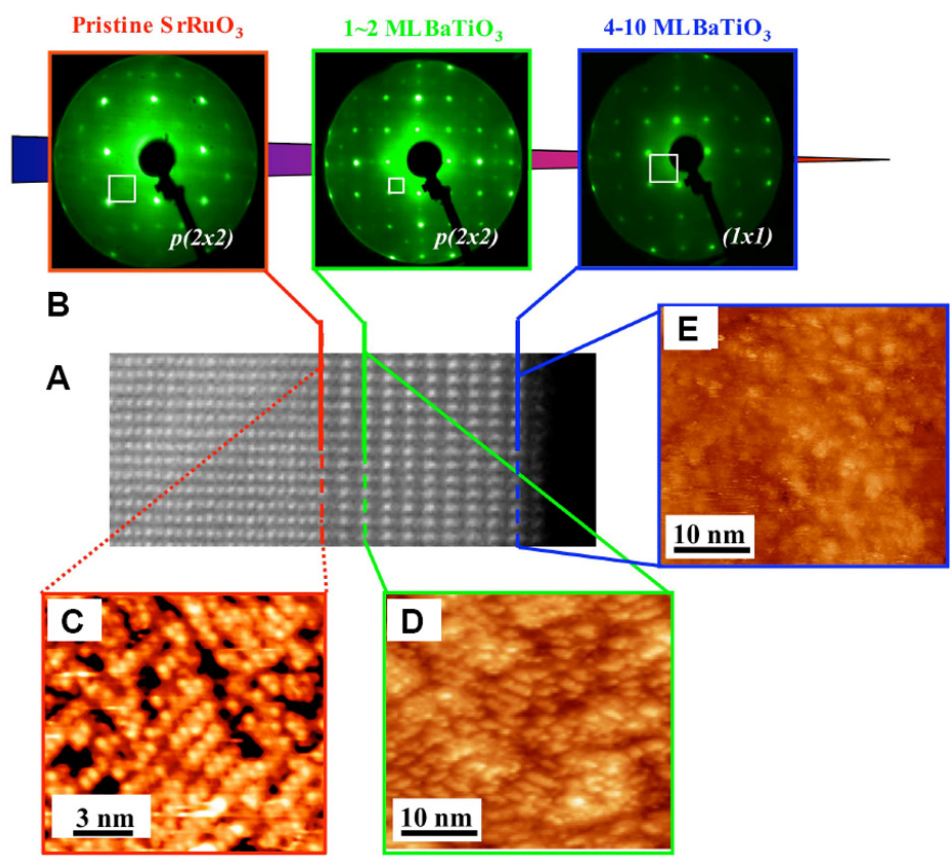

**Figure 5**